\documentclass[lettersize,conference]{IEEEtran}
\IEEEoverridecommandlockouts
\usepackage{cite}
\usepackage{amsmath,amssymb,amsfonts}
\usepackage{algorithmic}
\usepackage{graphicx}
\usepackage{textcomp}
\usepackage{xcolor}
\usepackage[inline]{enumitem}
\usepackage{subcaption}
\usepackage{easyReview}
\usepackage{array}
\usepackage{multirow}
\usepackage{float}
\usepackage{url}

\def\BibTeX{{\rm B\kern-.05em{\sc i\kern-.025em b}\kern-.08em
    T\kern-.1667em\lower.7ex\hbox{E}\kern-.125emX}}

\author{IEEE Publication Technology,~\IEEEmembership{Staff,~IEEE,}
\thanks{This paper was produced by the IEEE Publication Technology Group. They are in Piscataway, NJ.}
\thanks{Manuscript received April 19, 2021; revised August 16, 2021.}}

\usepackage[acronym,nopostdot,  style=super, nonumberlist, toc]{glossaries}
\newacronym[plural=HITLs,firstplural=human-in-the-loop (HITL)]{HITL}{HITL}{human-in-the-loop}
\newacronym[plural=CPSs,firstplural=cyber-physical systems (CPS)]{CPS}{CPS}{cyber-physical system}
\newacronym[]{TSN}{TSN}{time sensitive networking}
\newacronym[plural=CPSs]{UE}{UE}{user equipment}
\newacronym{HARQ}{HARQ}{Hybrid-ARQ}
\newacronym{MCS}{MCS}{modulation coding scheme}
\newacronym{MLP}{MLP}{multi-layer perceptron}
\newacronym{SDR}{SDR}{software-defined radio}
\newacronym{OAI}{OAI}{OpenAirInterface}
\newacronym{SINR}{SINR}{signal to interference and noise ratio}
\newacronym{CDF}{CDF}{cumulative distribution function}
\newacronym{RTT}{RTT}{round-trip time}
\newacronym{PTP}{PTP}{precision time protocol}
\newacronym{TDD}{TDD}{time division duplex}
\newacronym[plural=RANs,firstplural=radio access networks (RAN)]{RAN}{RAN}{radio access network}
\newacronym{COTS}{COTS}{commercial off-the-shelf}
\newacronym[plural=PDFs,firstplural=probability density functions (PDF)]{PDF}{PDF}{probability density function}

\usepackage{amsmath,amsfonts} 
\usepackage{graphicx}

\DeclareMathOperator*{\argmin}{arg\,min}

\begin{document}

\title{EDAF: An End-to-End Delay Analytics Framework for 5G-and-Beyond Networks}

\author{
    \IEEEauthorblockN{Samie Mostafavi\IEEEauthorrefmark{1}, Marius Tillner\IEEEauthorrefmark{2}, Gourav Prateek Sharma\IEEEauthorrefmark{3},
    James Gross\IEEEauthorrefmark{4}}
    \IEEEauthorblockA{\text{Silicon Austria Labs\IEEEauthorrefmark{2}} \text{KTH Royal Institute of Technology\IEEEauthorrefmark{1}\IEEEauthorrefmark{3}\IEEEauthorrefmark{4}}  \\
    ssmos@kth.se\IEEEauthorrefmark{1}, marius.tillner@silicon-austria.com\IEEEauthorrefmark{2}, gpsharma@kth.se\IEEEauthorrefmark{3}, jamesgr@kth.se\IEEEauthorrefmark{4}}
}

\maketitle

\thispagestyle{plain}
\pagestyle{plain}

\begin{abstract}

Supporting applications in emerging domains like cyber-physical systems and human-in-the-loop scenarios typically requires adherence to strict end-to-end delay guarantees.
Contributions of many tandem processes unfolding layer by layer within the wireless network result in violations of delay constraints, thereby severely degrading application performance.
Meeting the application’s stringent requirements necessitates coordinated optimization of the end-to-end delay by fine-tuning all contributing processes.
To achieve this task, we designed and implemented EDAF, a framework to decompose packets' end-to-end delays and determine each component's significance for 5G network.
We showcase EDAF on OpenAirInterface 5G uplink, modified to report timestamps across the data plane.
By applying the obtained insights, we optimized end-to-end uplink delay by eliminating segmentation and frame-alignment delays, decreasing average delay from 12ms to 4ms.


\end{abstract}

\begin{IEEEkeywords}
End-to-end delay, 5G, OpenAirInterface
\end{IEEEkeywords}

\section{Introduction}

In the realm of modern communication systems, a plethora of time-critical applications are emerging, e.g., smart manufacturing, Extended Reality (XR) and exoskeletons \cite{D11-d6g}. 
These applications necessitate that communication systems not only support low latency but with a certain level of reliability.
The performance of these applications is highly susceptible to variations in end-to-end packet delays, especially instances where delays deviate from the norm, as such outliers degrade the application's performance \cite{sharma2023deterministic}.
The end-to-end delay is a product of many tandem processes unfolding layer by layer within the wireless network, each exhibiting stochastic behavior \cite{D21-d6g}.
For instance, inside 5G's Radio Access Network (RAN), numerous processes, each serving different purposes, contribute to the end-to-end delay.
Radio Link Control (RLC) queuing, resource scheduling, packet segmentation, Hybrid Automatic Repeat reQuest (HARQ), and more are among these processes.


Meeting the application's stringent requirements necessitates coordinated optimization of the end-to-end delay by fine-tuning all contributing processes.
An effective approach involves firstly, conducting end-to-end delay measurements, and secondly, decomposing them into corresponding components while assessing the significance of each component.
This will enable further efficient and practical delay optimization.

To address this challenge, it is imperative to identify the sources of delay within the network under study and formulate an end-to-end decomposition model where each delay component represents a distinct delay source.
Next, according to the devised model, we record timestamps coupled with relevant information throughout the traverse of packets across the network.
The resulting decomposition supplies essential inputs for the subsequent delay optimization process.

In this work, we design and implement an end-to-end delay analytics framework, EDAF, to tackle these challenges and facilitate delay optimization for enhancing the performance of time-critical applications in 5G/5G-Advanced networks.

\subsection{Related Works}
3GPP has already introduced specifications for the Network Data Analytics Function (NWDAF) in 5G. NWDAF is an NF primarily responsible for collecting large amounts of data from  UEs and 5G Network Functions such as Access Management Function (AMF) and Session Management Function (SMF). This enables NWDAF to perform analytics and leverage, e.g., Machine Learning (ML) algorithms for automating network operations to optimize resource allocations. However, supporting time-critical applications would require advanced predictive analytics that can be translated into real-time operational intelligence. As these applications can have extreme requirements (e.g., 10 ms One-way Delay (OWD) with 99.999\% reliability), NWDAF needs to be enhanced \cite{pateromichelakis2019end,larrabeiti2023toward}. 
This requires, first, a fine-grained per-packet data collection mechanism that can track components of end-to-end latencies along with corresponding network states (e.g., Modulation and Coding Scheme (MCS) index, HARQ rounds) for these packets \cite{3gpp.38.314}. Second, certain analytics are required that can process this data to manage extreme latency and reliability requirements. Authors in \cite{ronteix2021latseq}, have proposed Latseq, a tool for the analysis of packet delays only in OpenAirInterface LTE basestation, not end-to-end. In addition, the majority of predictive analytics are too coarse-grained and thus are not suited for providing operations to guarantee tail latency \cite{sevgican2020intelligent}.
The data collection mechanisms discussed above are incapable of fine-grained data required for predictive analysis. For instance, LatSeq is capable of only tracing packet traversal through different layers inside the OpenAirInterface LTE stack at the basestation which might not be suitable for comprehensive end-to-end delay decomposition between the two application endpoints. A key requirement for implementing and evaluating this framework is an adaptable experimentation platform capable of supporting comprehensive end-to-end experiments. ExPECA, a testbed designed for wireless communication and edge-computing research, can support repeatable experimentation by leveraging the flexibility of software-defined radios and open-source 5G implementations such as OpenAirInterface5G \cite{mostafavi2023expeca, KALTENBERGER2020107284}.

\subsection{Contributions}

\begin{enumerate}
\item We introduce EDAF, a novel framework which is designed to conduct end-to-end delay decomposition and analysis by a) inserting time measurement points across the 5G protocol stack and application end points; and b) aggregating the measurements from all locations at the EDAF server.
\item The implementation of EDAF is demonstrated on the Openairinterface 5G network (uplink only) using ExPECA testbed's software-defined radios.
\item By leveraging EDAF on OpenAirInterface 5G uplink, we optimized its end-to-end delay by eliminating segmentation and frame-alignment delays, decreasing the average delay from 12ms to 4ms.
\end{enumerate}

\section{Problem Description}


In this work we aim to establish a methodology for decomposing end-to-end delays, allowing for separate measurement and optimization of each component.
Based on the measurements, inform the end-to-end delay violation probability for a given delay target and the degree of contribution of each component to the violations.
Such decomposing into $M$ additive components can be modelled as
\begin{align}
    Y_n(\boldsymbol{\theta}) = \sum_{j=1}^{M} Y_{n}^{(j)}(\theta_{j}),
\end{align}
where $Y_n$ denotes the end-to-end delay of packet $n$, $\boldsymbol{\theta} \in \Theta$ is the set of influential parameters, and $\theta_j$ is the subset of parameters affecting the $j$th delay component: $Y_n^{(j)}$.
Furthermore, it is crucial to identify the degree that each component contributes to delay violations for a given delay target $\tau$ as
\begin{align}
    \mathbb{E}\left[ \frac{Y_n^{(j)}}{Y_n} \mid Y_n > \tau, \boldsymbol{\theta} \right].
\end{align}
By determining the significance of each component's contribution, we can sort them accordingly and apply latency minimization iteratively, focusing on the most impactful components first. 
This targeted and efficient approach ensures that the delay minimization process is directed towards the most influential factors.


\section{Delay Decomposition Model}
\label{sec:delay_decomposition}

\begin{figure}[t]
    \centering
    \includegraphics[width=0.99\linewidth]{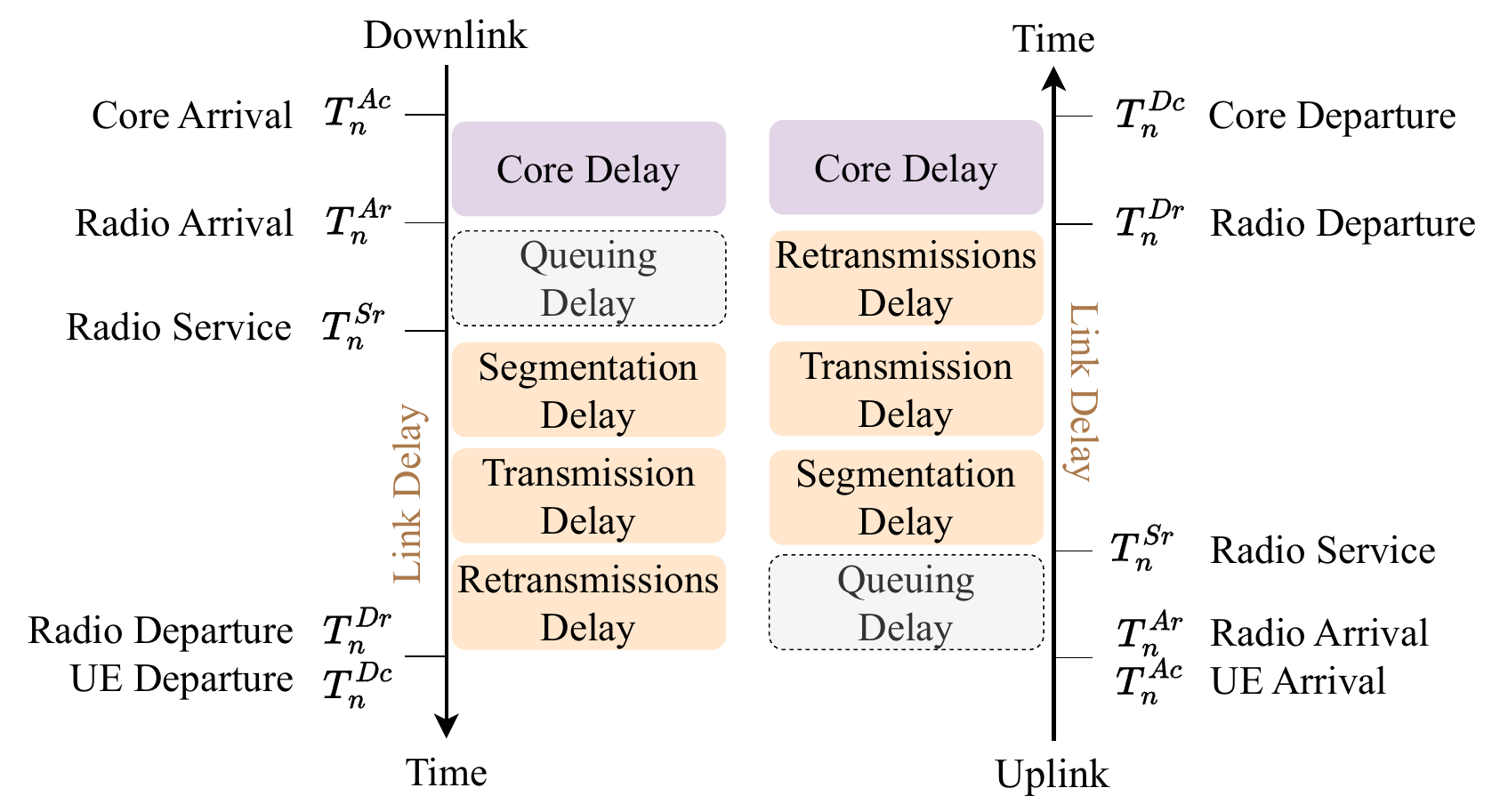}
    \caption{Packet's end-to-end delay components in EDAF}\label{fig:delay-components}
\end{figure}

In this section, we decompose the delay of a packet traveling through the network either on the uplink or downlink, as a sum of three major components as shown in Figure \ref{fig:delay-components}: core delay, queuing delay, and link delay.
This separation firstly is a result of the well-known division of mobile networks into two main domains: core network and RAN. A more comprehensive breakdown of end-to-end packet delay for 5G is presented in \cite{D21-d6g}.
The core delay component can become influential in the cases where the network's gateway is multiple hops away from the RAN.
It can be measured by timestamping the packets entering and exiting the N3 interface between the RAN and UPF.
On the uplink, we refer to these timestamps as radio departure and core departure, on the downlink, UE arrival and radio arrival \cite{5g_ran_2021,e2e_v2x}.

Second, we resort to a queuing model for the RAN delay and this results in another split into queuing delay and link delay.
Due to the shared and stochastic capacity of the wireless channel, IP packets, or parts of them, can be buffered upon arrival since the wireless link is busy transmitting the previous packets.
Also, packets are queued when the radio waits for a transmission opportunity i.e. frame-alignment delay or waits to receive a transmission grant from the scheduler.
Such buffering mechanisms are implemented at various levels of the network.
We recognized that the closest queue to the wireless link, which is an RLC queue in 5G, contributes the most to the end-to-end delay and its variations.
Therefore, we orient our approach and feature a singular queue in the delay model.


In terms of notation, we represent all timestamps and delays as random variables, denoted by $T$ and $Y$, respectively, and utilize subscripts to indicate the packet index, ranging from $n$ to a total of $N$ packets. 
Additionally, superscripts are employed to differentiate between types of timestamps. 
For instance, $T^{Ac}_{n}$ signifies the arrival timestamp of packet $n$ at the core level, as depicted in Figure \ref{fig:delay-components}. 

We denote the queuing delay of packet $n$ by $Y_{n}^{Q}$ and measure it by subtracting its radio arrival time and radio service time: $Y_{n}^{Q} = T_{n}^{Sr} - T_{n}^{Ar}$.
Radio arrival time refers to the time when the packet enters the radio link buffer or RLC queue.
Radio service time represents the time when the MAC layer starts consuming the packet for transmission.

It is crucial to analyze the process in which the radio service time of a packet, $T^{Sr}_{n}$, is determined as it is a delay variation source.
When no packets are left to be served in front, the service time starts some time i.e., preparation time, earlier than the arrival of the granted slot.
Therefore, the queuing time could be due to the division of time slots to uplink or downlink in a repeating pattern in TDD networks.
Moreover, in grant-based scheduling, the waiting time until a grant is received for the transmission is included in the queuing delay.


\begin{figure}[t]
    \centering
    \includegraphics[width=0.9\linewidth]{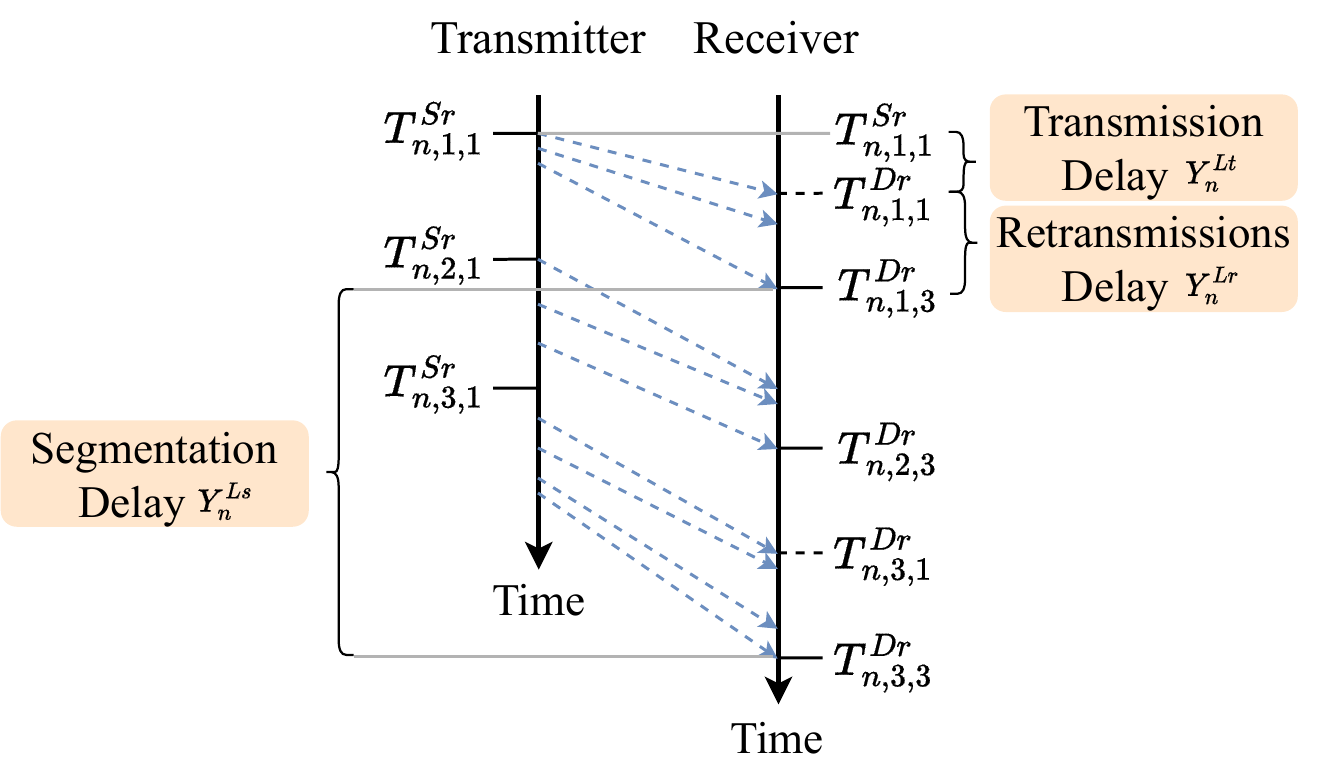}
    \caption{Link delay components in EDAF. Dashed arrows indicate HARQ attempts.}\label{fig:link-components}
\end{figure}

Next, we introduce the radio link delay $Y_{n}^{L}$ which focuses on the period that starts after queuing, i.e., radio service time, until the packet is reassembled successfully at the RLC layer of the receiver, i.e., radio departure time: $Y_{n}^{L} = T_{n}^{Dr} - T_{n}^{Sr}$.

In the case of timestamps within the radio link, the subscripts, in addition to the packet index, refer to a HARQ attempt denoted by $l$ out of a total of $L$ attempts and a segment denoted by $m$ out of a total of $M$ segments.
An illustrative example is given by $T^{Dr}_{n,m,l}$, representing the departure timestamp of packet $n$ in the radio layer, specifically for the $m$-th segment and the $l$-th HARQ attempt as shown in Figure $\ref{fig:link-components}$ \cite{DAHLMAN202179}.
We describe the three link latency components as follows:

\subsubsection{Segmentation Delay $Y_{n}^{Ls}$}
Upon the packet's service time, a specific allocation of Physical Resource Blocks (PRBs) within a time slot is granted for transmission. 
Simultaneously, the Modulation and Coding Scheme (MCS) index is derived from the measured channel quality to minimize the Block Error Rate (BLER) and maximize spectral efficiency. 
The combination of allocated PRBs and the MCS index result in the number of bytes that can be consumed from the queue, referred to as the Transport Block Size (TBS). 
TBS may be smaller than the original packet size. 
In such cases, the packet undergoes segmentation, and the segments are sequentially transmitted in the upcoming transmission slots. 


\subsubsection{Retransmissions Delay $Y_{n}^{Lr}$}
Hybrid Automatic Repeat Request (HARQ) is used in most wireless communication schemes to ensure reliability by retransmitting lost or corrupted packets. 
It works by combining Forward Error Correction (FEC) with Automatic Repeat Request (ARQ) protocols, allowing for error correction.
However, it introduces additional retransmission delay for some packets.
This component of delay which is related to the MCS index, the channel conditions, and the resulting number of retransmissions is defined as the time difference between the first to the last transmission attempt of one segment of the packet.

\subsubsection{Transmission Delay $Y_{n}^{Lt}$}
The delay of one attempt on transmission of a segment by the radio link is denoted by transmission delay.
In other words, it is the time it takes for MAC to encode, modulate, and send a segment of the packet and receive, demodulate, and decode it at the receiver, regardless of the HARQ decoding result.

Precisely decomposing a link delay into the aforementioned three components poses a challenge, primarily due to the assignment of PDU segments to multiple parallel HARQ processes, designed for enhanced throughput and efficiency, as depicted in Figure \ref{fig:link-components}.
Each segment entails distinct transmission and retransmission delays, and their potential overlap arises when they belong to different segments owing to HARQ pipelining. 
In response to this complexity, we propose a systematic decomposition of link delay for each packet's traverse. 
Initially, we identify the segment with the maximum segment service delay, denoted as $m^{*}$. 
Subsequently, we decompose its service delay, saving them as the transmission and retransmission delays of the packet: $Y_{n}^{Lt} = T_{n,m^{*},1}^{Dr}-T_{n,m^{*},1}^{Sr}$ and $Y_{n}^{Lr} = T_{n,m^{*},L}^{Dr}-T_{n,m^{*},1}^{Dr}$. 
The remaining link delay is then considered as the segmentation delay, calculated as $Y_{n}^{Ls} = Y_{n}^{L} -Y_{n}^{Lt} -Y_{n}^{Lr}$. 
We summarize the additive components as $Y_{n} = Y_{n}^{C} + Y_{n}^{Q} + Y_{n}^{L}$ and $
Y_{n}^{L} = Y_{n}^{Lt} + Y_{n}^{Ls} + Y_{n}^{Lr}$.

\begin{table}[t]
\centering
    \caption{Features to collect alongside the timestamps}
        \begin{tabular}{|l|l|}
            \hline
            \multicolumn{1}{|l|}{\textbf{Delay component}} & \multicolumn{1}{l|}{\textbf{Information to Collect}} \\
            \hline
            \multirow{3}{*}{Queueing delay} 
                 & \multicolumn{1}{l|}{Arrival Time} 
                 \\\cline{2-2}
                 & \multicolumn{1}{l|}{Queue length (bytes)} 
                 \\\cline{2-2}
                 & \multicolumn{1}{l|}{First scheduled slot} 
                 \\\cline{2-2}
            \hline
            \multirow{3}{*}{Segmentation delay} 
                & \multicolumn{1}{l|}{Packet size (bytes)} 
                \\\cline{2-2}                     
                & \multicolumn{1}{l|}{TBS (bytes)} 
                \\\cline{2-2}
                & \multicolumn{1}{l|}{Segments scheduled slots}
                \\\cline{2-2}
            \hline
            \multirow{4}{*}{Retransmissions delay}
                & \multicolumn{1}{l|}{Number of Retransmissions} 
                \\\cline{2-2}
                & \multicolumn{1}{l|}{Retransmission slots}
                \\\cline{2-2}
                & \multicolumn{1}{l|}{MCS index} 
                \\\cline{2-2}
                & \multicolumn{1}{l|}{ Channel quality indicators} 
                \\\cline{2-2}
            \hline
        \end{tabular}
    \label{tab:features}
\end{table}

Furthermore, as described in each delay component, we recognize that the parameters listed in Table \ref{tab:features} are critical to collect and store, for every packet.
This information is considered critical to collect since they have the main influence on each delay component.
In the next section, we dive into the intricacies of the EDAF implementation.

\section{EDAF Implementation}
\label{sec:impl}

\begin{figure}[t]
    \centering
    \includegraphics[width=0.99\linewidth]{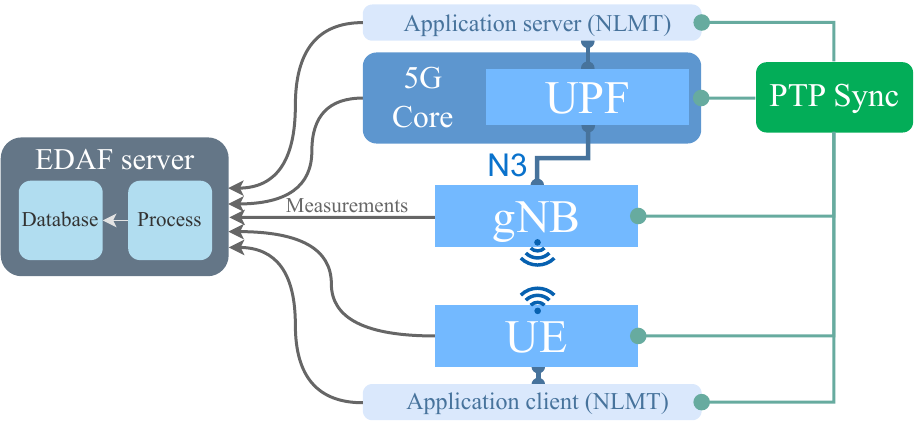}
    \caption{EDAF implementation setup}\label{fig:setup}
\end{figure}

The implementation of EDAF serves two primary purposes: 1) to measure and store the end-to-end delay and its components for each packet; and 2) to process and analyze the collected data, presenting inferred outcomes for integration into end-to-end delay optimization.
The first is achieved by recording timestamps, starting from the application's client node, across the 5G network, and ending at the application's server node, to record the packets' traverse end-to-end.
In practice, we made specific design choices in the EDAF implementation, summarized as follows:

\textbf{OpenAirInterface:} To insert measurement points across layers of the 5G network, spanning from the application layer down to the MAC layer's HARQ process, and consequently access the produced data with minimal effort lead us to opt for OpenAirInterface (OAI) SDR-based 5G implementation to host EDAF.
OAI's open-source code is adaptable to different use cases and new functionality can be implemented ~\cite{KALTENBERGER2020107284}.

\textbf{LatSeq:} Data collection is crucial to be conducted with minimal impact on the user flow, adhering to an approach that does not alter the packet or its processing.
Hence, we chose the LatSeq project, designed for extracting timestamped information with minimal overhead from various layers within OAI, as the main data collection tool ~\cite{ronteix2021latseq}.

\textbf{PTP synced hosts:} Achieving time synchronicity among entities conducting time measurements was crucial for interpreting time differences.
All nodes were equipped with hardware timestamping-capable network interface cards (NICs) to support Precision Time Protocol (PTP) for clock synchronization on an out-of-band network.

\textbf{Microservices:} Real-time data collection and processing are crucial for the delay optimization process. 
To address this need, we opted for a microservices architecture in the development of EDAF, as opposed to the file-based data exchange method utilized in LatSeq. 
EDAF is encapsulated as a Docker container, utilizing network connections for data collection from all nodes.

\textbf{Time series database:} Due to the probabilistic analysis of delay in EDAF, a substantial amount of data must be efficiently stored for time-based queries. Therefore, EDAF includes InfluxDB, a time-series based database, to store measurements and respond to queries efficiently.

The EDAF implementation in this study is limited to the 5G uplink. 
In this part, we begin by outlining the placement of measurement points within the UE RAN stack—specifically, as the packet traverses PDCP, RLC, MAC, and subsequently a HARQ process. 
Within the PDCP layer, timestamps are recorded when a packet enters and exits this layer, accompanied by the PDCP sequence number (SN). 
Transitioning to the RLC layer, timestamps are registered when a packet enters RLC data buffers or when the MAC layer initiates the extraction of the first segment. 
These timestamps serve as indicators of queuing time for the packets. 
In post-processing, the unique SN for each IP packet distinguishes its timestamps from others. 
However, beyond this point, SN is no longer extractable as we now deal with chunks of bytes. 
To address this, we save the memory location and size of the segments when taken by the MAC layer. 
Concurrently, in all active HARQ processes, timestamps are measured when any transmission attempt starts, capturing information such as the source memory location, HARQ ID, assigned MCS index, frame number, slot number, and PRBs. 
In post-processing, we track the segments of a packet by tracing their memory location down to the HARQ process.

Proceeding to the gNB, we initiate from the lowest layer. 
Similarly, in the gNB, all HARQ processes timestamp decoding attempts, noting their frame number, slot number, and HARQ ID.
This information pertains to the same segment in UE, enabling the establishment of a connection between gNB and UE traces and complete the end-to-end analysis.
While the remaining layers in the gNB operate similarly, detailed coverage is constrained due to space limitations.

EDAF is available as open-source software, and can be accessed on GitHub \footnote{https://github.com/samiemostafavi/edaf}. 
The repository provides comprehensive instructions on running EDAF, including the setup for the modified OAI 5G network with SDRs.

\section{Experiments and Numerical Analysis}
\label{sec:numerics}

\begin{figure}[t]
    \centering
    \includegraphics[width=0.99\linewidth]{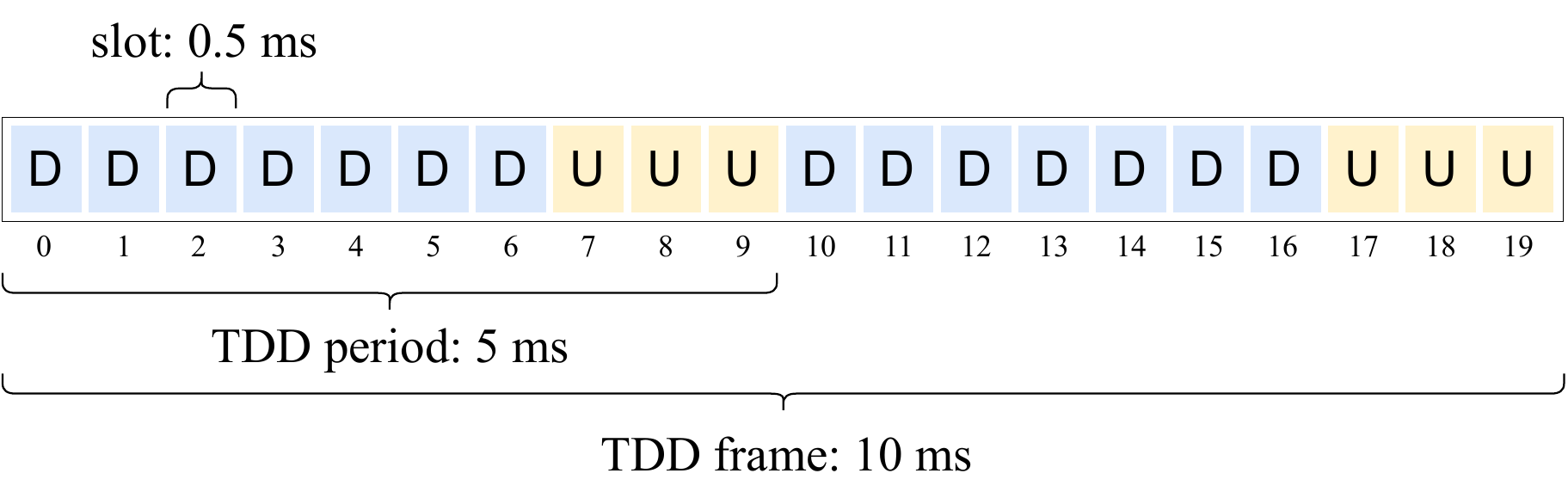}
    \caption{Experiments TDD frame format, "D" indicates downlink slots and "U" is uplink.}\label{fig:tddpattern}
\end{figure}

In this section, we showcase an end-to-end (e2e) delay optimization task using EDAF-generated insights for a packet flow traversing over the uplink of our SDR-based 5G network.
This particular traffic is characterized by the periodic transmission of small packets, demanding e2e delay target $\tau$ with a reliability level of $\epsilon$.
The task of managing the delay requirements mandates choosing parameter set $\boldsymbol{\theta}$ in such a manner that $\mathbf{P}[ Y_n(\boldsymbol{\theta}) > \tau \mid \boldsymbol{\theta} ] < \epsilon$.
We assumed two different e2e delay requirements to examine: ($\tau_1 = 5 \text{ms}$, $\epsilon_1 = 10^{-2}$) and ($\tau_2 = 15 \text{ms}$, $\epsilon_1 = 10^{-4}$).
As for the controllable parameters, the number of PRBs in the uplink grant and the packet arrival time offset are considered.

\begin{figure}[t]
	\centering
	\begin{subfigure}{0.85\linewidth}
		\includegraphics[width=\linewidth]{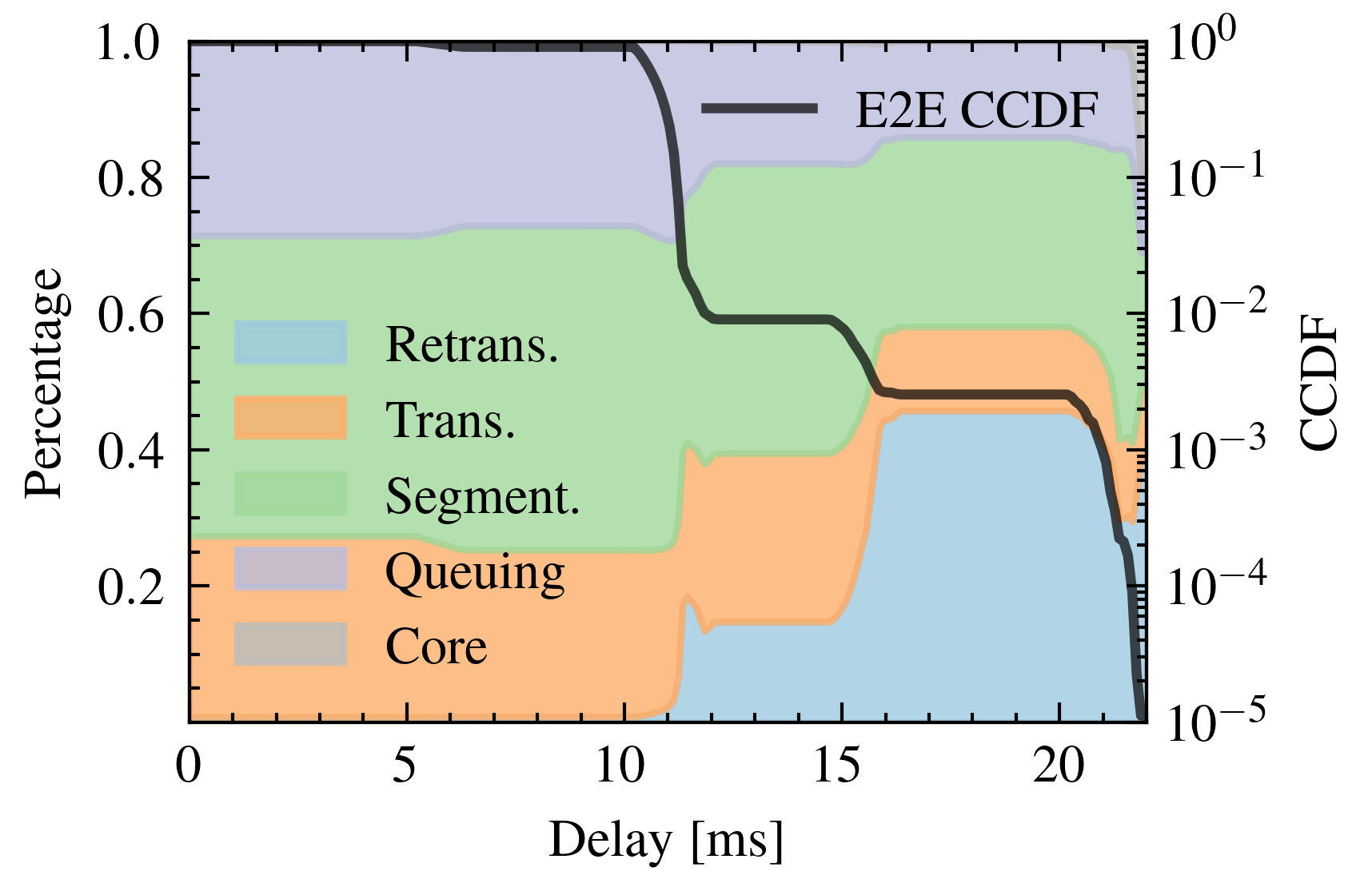}
		\caption{5 PRBs $\rightarrow$ TBS of 396 bytes}
            \label{fig:decompose-e2e-a}
	\end{subfigure}
	\begin{subfigure}{0.85\linewidth}
		\includegraphics[width=\linewidth]{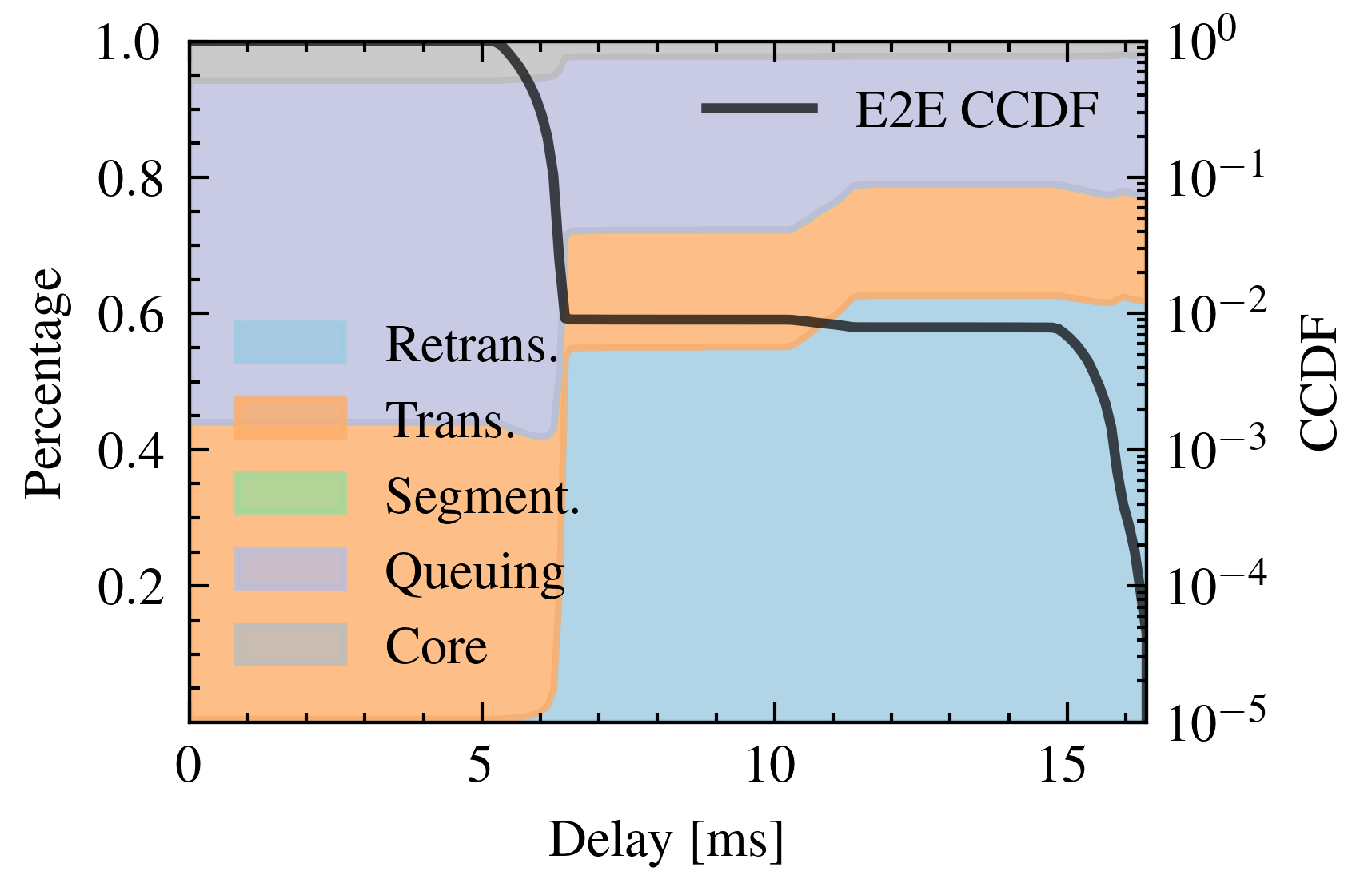}
		\caption{10 PRBs $\rightarrow$ TBS of 792 bytes}
            \label{fig:decompose-e2e-b}
	\end{subfigure}
        \begin{subfigure}{0.85\linewidth}
		\includegraphics[width=\linewidth]{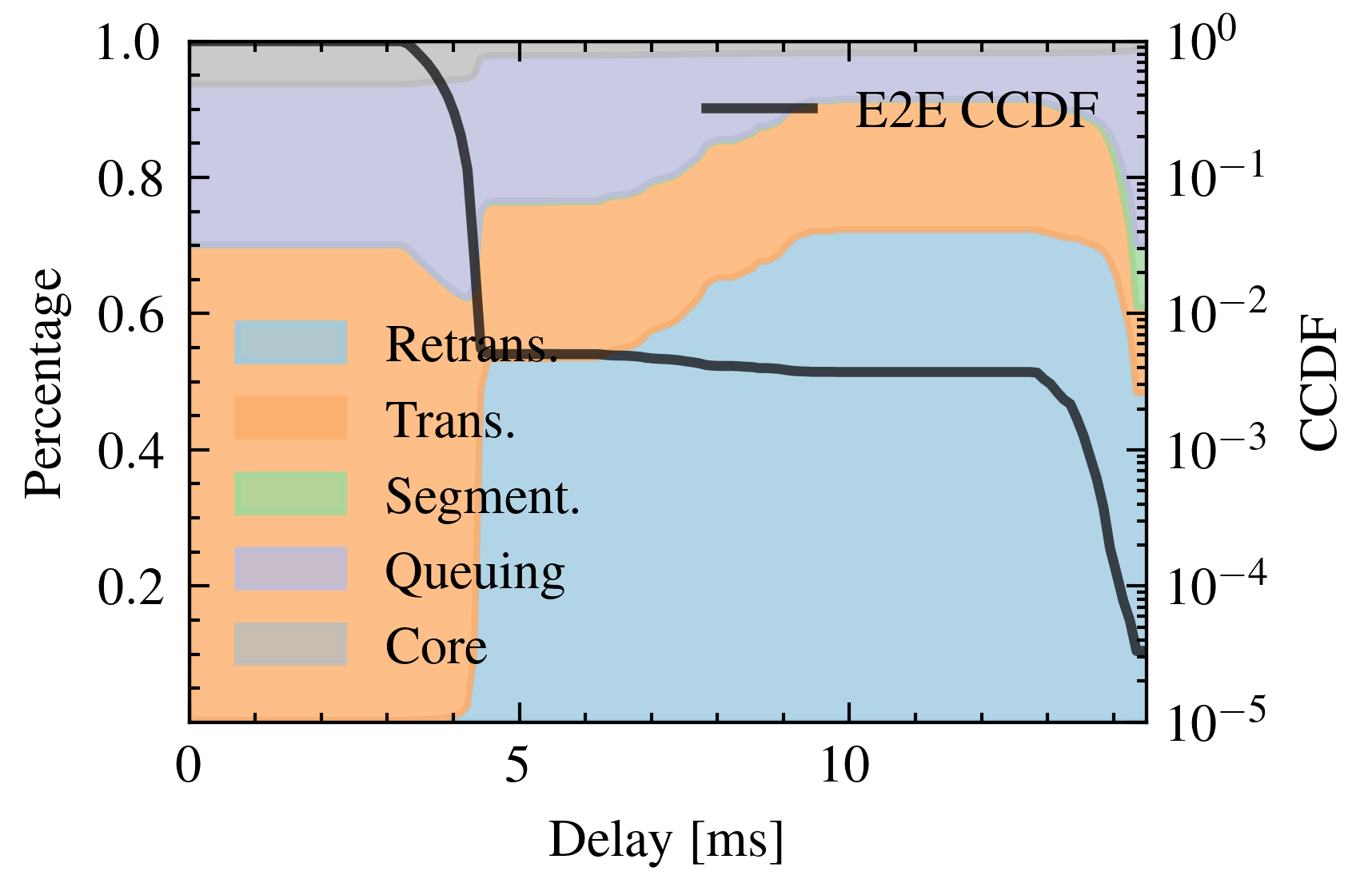}
		\caption{10 PRBs + optimized packet arrival times}
            \label{fig:decompose-e2e-c}
	\end{subfigure}
	\caption{EDAF e2e CCDF and decomposition in experiments feature a fixed MCS index of 23 and 500-byte packets.}
	\label{fig:decompose-e2e}
\end{figure}

\begin{figure}
	\centering
	\begin{subfigure}{1\linewidth}
		\includegraphics[width=\linewidth]{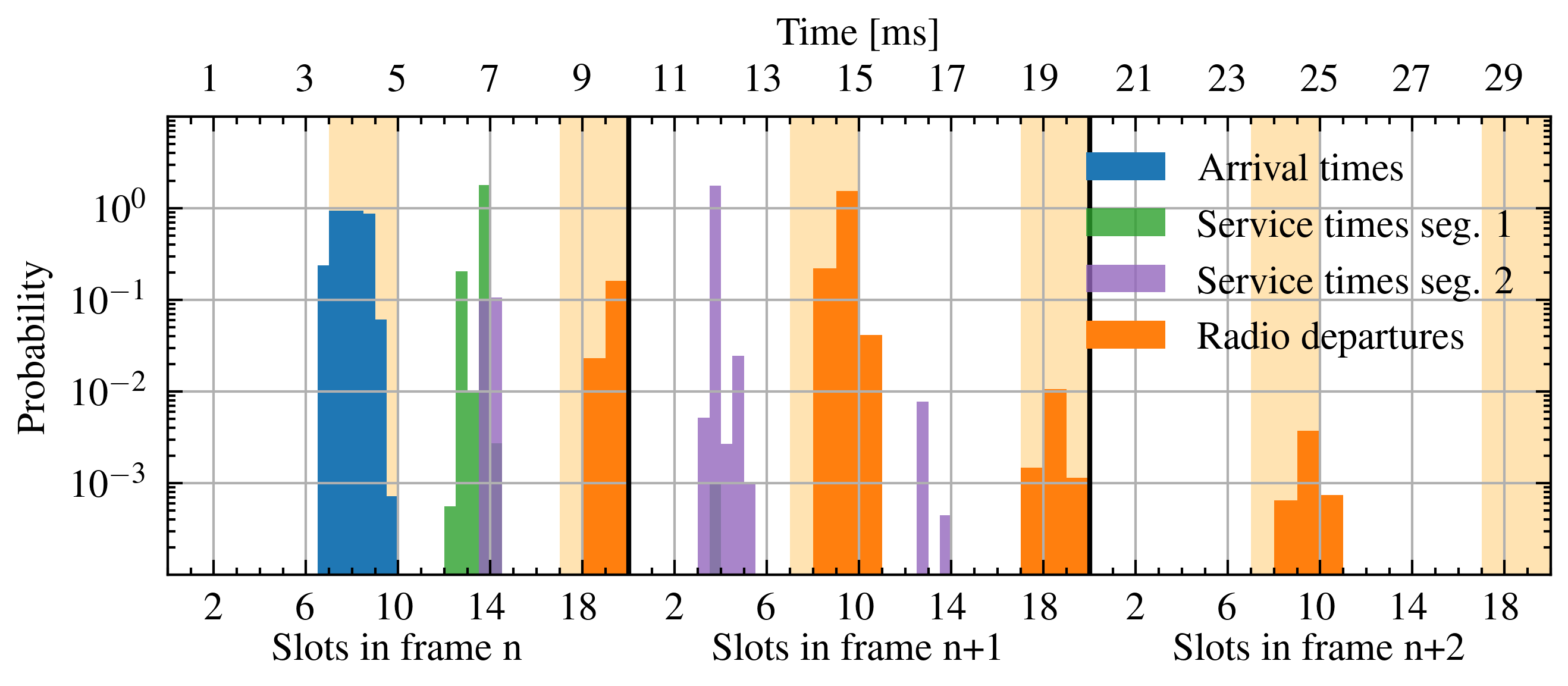}
		\caption{5 PRBs $\rightarrow$ TBS of 396 bytes}
            \label{fig:time-e2e-a}
	\end{subfigure}
	\begin{subfigure}{1\linewidth}
		\includegraphics[width=\linewidth]{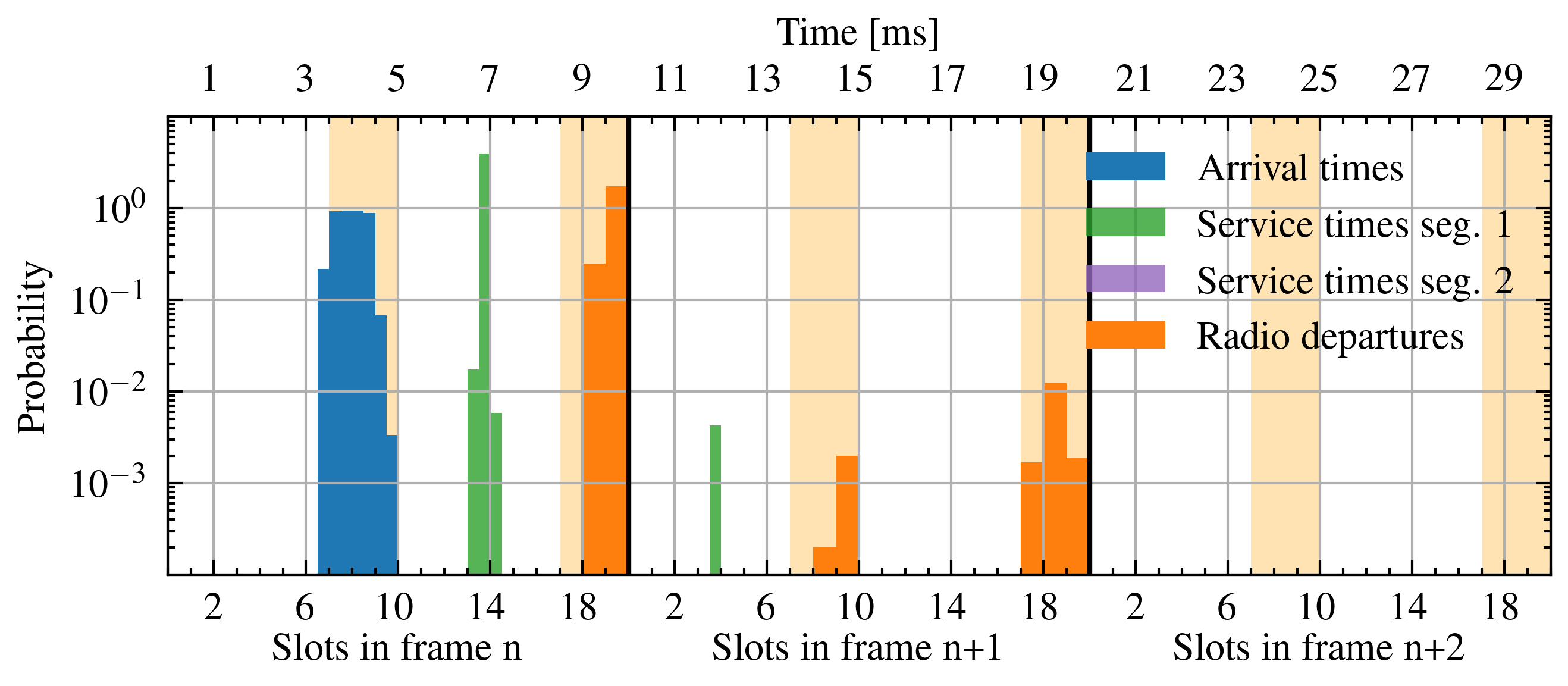}
		\caption{10 PRBs $\rightarrow$ TBS of 792 bytes}
            \label{fig:time-e2e-b}
	\end{subfigure}
        \begin{subfigure}{1\linewidth}
		\includegraphics[width=\linewidth]{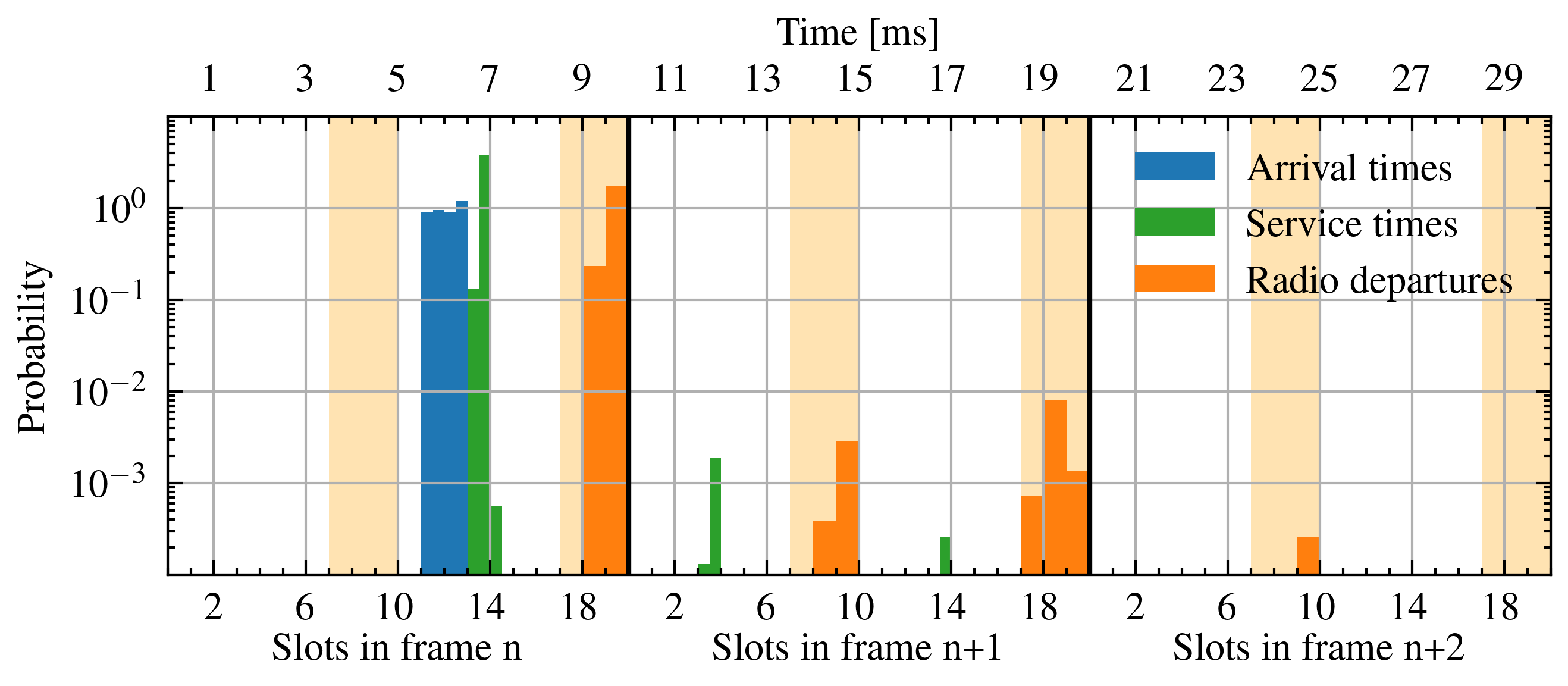}
		\caption{10 PRBs + optimized packet arrival times}
            \label{fig:time-e2e-c}
	\end{subfigure}
	\caption{Histograms of service times and radio departure times, relative to the arrival frame of the packets in Figure \ref{fig:decompose-e2e}.}
	\label{fig:time-e2e}
\end{figure}

For each experiment that lasted for about 20 minutes, we used Network Latency Measurement Tool (NLMT) developed by us to generate periodic UDP packets, capable of reporting packets' sequence numbers and timestamps to the EDAF server.
Traffic packets each measuring 500 bytes, were sent at intervals of 10 milliseconds in the uplink direction.
In total each experiment produced 120.000 packets.
The OAI 5G network, operating in band 78 and employing TDD mode, featured 106 PRBs, occupying a 40 MHz bandwidth at 3.5 GHz.
The experiment's default TDD pattern, depicted in Figure \ref{fig:tddpattern}, plays a crucial role in comprehending and identifying latency causes within the network.
The 5G frames have a duration of 10 ms, resulting arrival of one packet per frame in the experiments.
The NLMT tool can adjust the packet arrival time relative to the start of the TDD frame with a precision of 1.5 ms, exemplified in Figure \ref{fig:time-e2e}. 
Blue histogram in this figure visualizes the distribution of packets' relative arrival time, set to 4 ms but extending across a range.



We start with Experiment (a) which serves as the baseline, adhering to the default configuration.
EDAF analysis is illustrated in Figures \ref{fig:decompose-e2e} and \ref{fig:time-e2e}, with their respective sub-figures corresponding to Experiments (a), (b), and (c).
Figure \ref{fig:decompose-e2e} displays the Complementary Cumulative Distribution Function (CCDF) for the measured e2e delays which is an estimate of the Delay Violation Probability (DVP) for different e2e delay targets and in case of sufficiently large samples considered accurate.
In addition, the left axis depicts the decomposed contribution percentage of the components to the cumulative e2e delay.
First, we assess the e2e DVP for a given $\tau$.
In Experiment (a), the 15 ms target has a DVP of $10^{-2}$ and 5 ms's DVP is almost 1.
Next, we examine the decomposition for each target delay.
For both targets, segmentation delay contributes $45\%$ and $40\%$ respectively, which is the most.

In Experiment (b), we eliminate segmentation delay by increasing the uplink grant PRBs to 10 from 5 which facilitates a TBS of 880 bytes, sufficient to accommodate an entire 531 bytes packet along with headers.
Experiment (a) with a TBS of 396 bytes necessitates packet segmentation, leading to a 5 ms e2e delay increment.
Figure \ref{fig:time-e2e} illustrates the probability distributions of radio arrival times, service times, and radio departure times relative to the start of the 5G frame that they arrive on.
Distributions are depicted on 3 consecutive frames to show the complete evolution of the packets' traverse in time.
In Figure \ref{fig:time-e2e-a}, the impact of segmentation is evident, where the service times of segment 2 occur 10 or 20 slots later than segment 1, resulting in a delay in the departure of the packet.
Conversely, in \ref{fig:time-e2e-b} and \ref{fig:time-e2e-c}, this occurrence is mitigated, leading to much-improved e2e delays.
However, in Experiment (b), we still observe the same DVPs for 15 and 5 ms target delays.
Hence, we continue the optimization.

\begin{figure}
\centering
\includegraphics[width=0.8\linewidth]{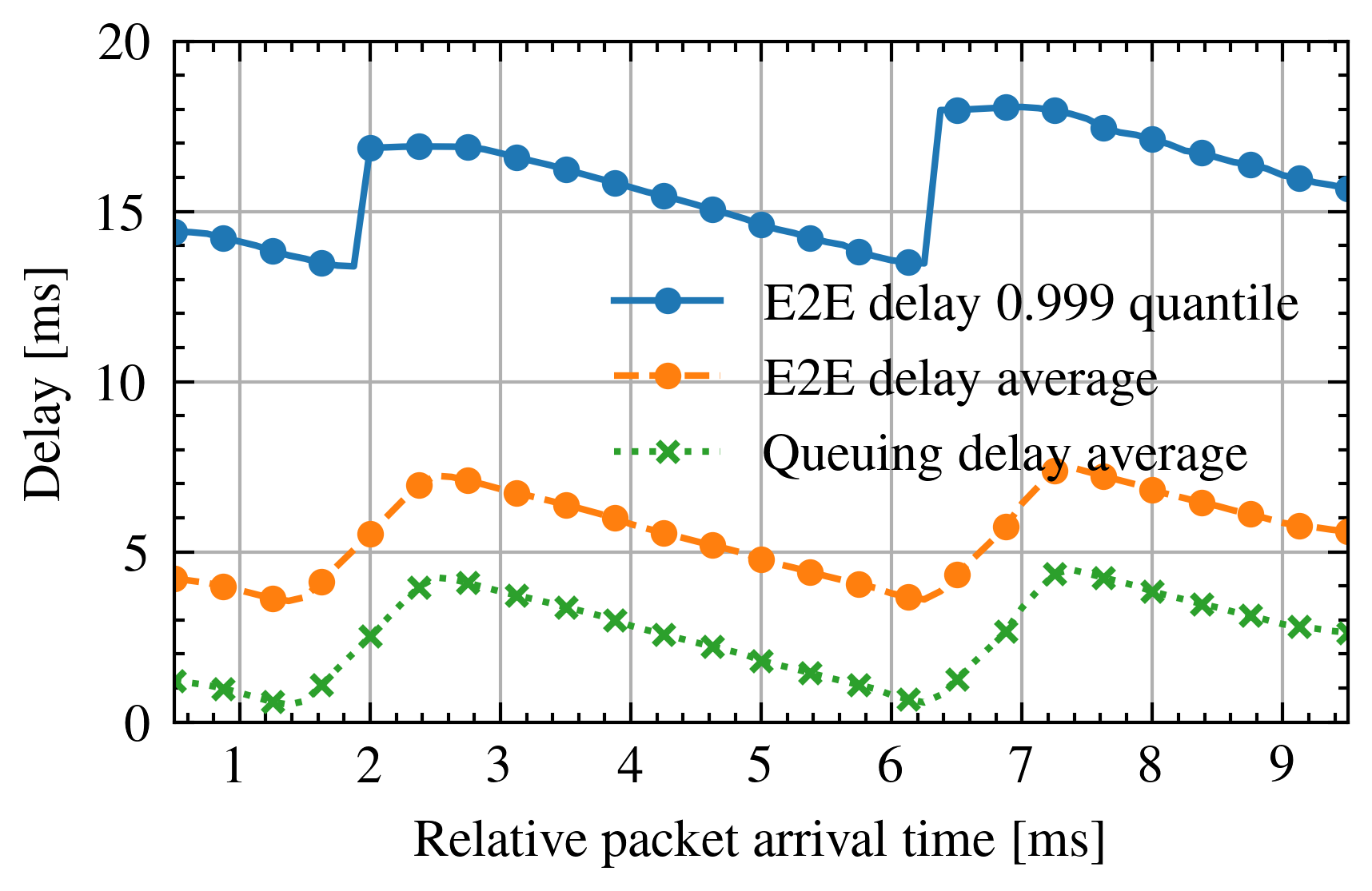}
    \caption{Minimizing end-to-end delay through optimizing arrival times offset relative to the 5G TDD period.}
\label{fig:queueplot}
\end{figure}
Experiment (c) eliminates segmentation delay and concurrently minimizes queuing delay by eliminating the frame-alignment delay.
The time gap observed in Figure \ref{fig:time-e2e-b}, spanning from arrival times in slots 7 to 10 to the first service times in slots 12 to 14, is attributed to the early arrival of the packets.
Therefore, this optimization is expected to reduce e2e delays by 3 ms.
For this task, we chose the arrival time offset denoted by $\theta$ to minimize the queuing delay formulated as
$\theta^{*} = \argmin_{\theta \in [0,10]}(Y_n^{Q})$.
The results of the search for the optimum $\theta$ are illustrated in Figure \ref{fig:queueplot} where the resulting e2e delay is also depicted.
As evidenced by the figure, our hypothesis was validated, with offsets of 1-2 and 6-7 yielding the minimum E2E delay. 
Conclusively, in this experiment, DVPs for both 5 ms and 15 ms are constrained to $10^{-2}$ and $10^{-4}$, meeting the application requirements and halting the delay minimization process.

Another notable observation is the decomposition for low e2e delays versus high e2e delays where retransmission delay starts to dominate, accounting for up to $50\%$ of the e2e delay.
This trend is further evident in all experiments, underscoring that the primary contributor to the extended tail in e2e delay is the infrequent yet impactful retransmissions.

\section{Conclusion}
\label{sec:conclusion}
In conclusion, this paper introduces EDAF, a novel tool that 1) modifies the 5G protocol stack by inserting numerous time measurement points; 2) requires the application to report packet timestamps; 3) aggregates all time measurements at a server to analyze the end-to-end delay; and 4) generates insights for delay optimization.
Conducted experiments on OpenAirInterface 5G network, highlight the potential benefits of EDAF, particularly in optimizing end-to-end packet delays.
We show the feasibility of a systematic optimization of the end-to-end delay, through EDAF's constant measurement and decomposition analysis.
Moreover, this tool can be useful for fellow researchers under a wide range of topics concerning latency in 5G and beyond 5G networks.
EDAF offers an end-to-end delay analysis setup in OpenAirInterface 5G which can facilitate testing the new low-latency functionalities.
For instance, 3GPP standardized Ultra Reliable Low Latency Communications (URLLC) features in 5G and there is an emerging interest in assessing their effectiveness in various conditions.
They can be developed on OpenAirInterface and utilize EDAF to test the deployment.
Moreover, using EDAF it is possible to quantify the probabilities of departure across various time slots.
Such insight can be harnessed for computing optimized wireless-friendly end-to-end schedules in order to realize the emerging goal of 5G-TSN integration \cite{D31-d6g}.

\section{Acknowledgements}
This work was supported by the European Commission through the H2020 project DETERMINISTIC6G (Grant Agreement no. 101096504).

\bibliographystyle{ieeetr}
\bibliography{refs.bib}

\appendices
\section{Demo}

\begin{figure}[t]
    \centering
    \includegraphics[width=0.99\linewidth]{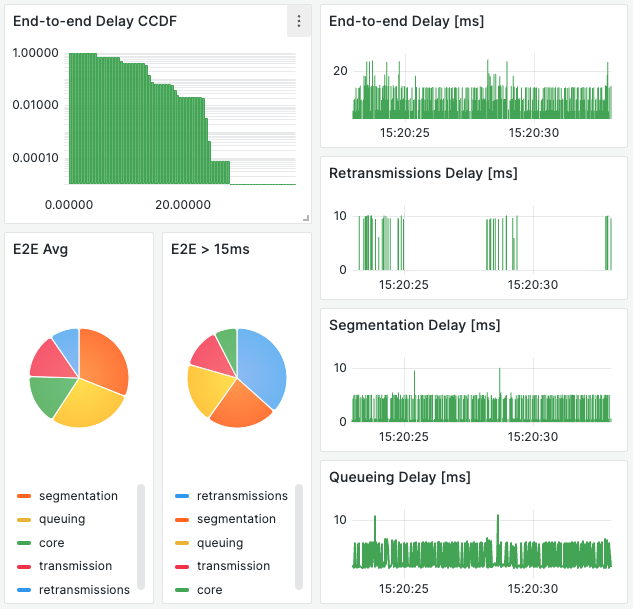}
    \caption{Grafana dashboard illustrating EDAF end-to-end delay analytics in real-time}\label{fig:demopanel}
\end{figure}

In this demo, we showcase EDAF implementation on OpenAirInterface 5G uplink, generating end-to-end delay analytics on a live online Grafana dashboard as shown in Figure \ref{fig:demopanel} for different network configurations.

We use the ExPECA testbed located at KTH university to run the experiments live as conducted for section \ref{sec:numerics}.
ExPECA testbed provides a highly reproducible environment for end-to-end wireless and edge computing experimentation, making it one of the first platforms to effectively address this issue.
The testbed is located in KTH R1 hall, an experimental facility 25 meters below ground level. 
This isolated location offers a controlled environment with minimal radio interference, making it ideal for running reproducible wireless experiments \cite{mostafavi2023expeca}.

We will employ 3 PTP-time synchronized servers to run the OAI core network, OAI gNB, and OAI UE, respectively.
As for the SDRs, we use 2 E320 USRPs for OAI gNB and OAI UE in this demonstration.
We will run a modified OpenAirInterface 5G stack which incorporates all measurement points and a modified version of LatSeq to forward the traces to the EDAF server running on the core network's host as mentioned in Section \ref{sec:impl}.
Then, we utilize our NLMT tool to generate periodic UDP traffic. 
It reports packets' timestamps to the EDAF server, and synchronize the transmission times on uplink with 5G's TDD frames.

As soon as the connection is established, EDAF processes the time measurements of all nodes in batches and populates InfluxDB.
We have deployed a Grafana server to illustrate the results by connecting to the EDAF server's InfluxDB.
It periodically retrieves the last 10-minute records and illustrates the e2e delay decomposition and CCDF in a web-based dashboard as shown in Figure \ref{fig:demopanel}.
The decomposition is shown by 2 Pie charts, first created from all packets, and second, from the end-to-end delays greater than 15ms to show the significance of retransmissions on the higher delays.
The CCDF plot and Pie charts are followed by time series plots of the components underneath the time series of the e2e delay to show their correlation during peaks.

The experiment will be done interactively, and members of the audience will be invited to propose different configurations regarding the 5G network or the traffic generation pattern.
For instance, it will be possible to test three configurations that optimize the e2e delay in an iterative manner:
\begin{enumerate}
    \item Baseline: allocating $X_1$ PRBs for the uplink grant.
    \item Eliminated segmentation delay: allocating $X_2 > X_1$ PRBs for the uplink grant.
    \item Eliminated segmentation and frame-alignment delays: allocating $X_2$ PRBs for the uplink grant and setting the optimum packet arrival offset.
\end{enumerate}
The number of PRBs $X_1$ and $X_2$ must be determined during the experiment since the MCS index of the connection and the packet size of the traffic are needed for its calculation.
It is expected to observe e2e delay improvements as we apply each new configuration.
We can repeat the experiments on different SDR nodes located at ExPECA R1 hall, causing different retransmission probabilities and consequently new e2e delay profiles.

In conclusion, this live demonstration of EDAF implementation on the OpenAirInterface 5G uplink, highlights the framework's capability to provide comprehensive end-to-end delay analytics for various network configurations in real time. 
Leveraging the ExPECA testbed's controlled environment, our experiment demonstrates the reproducibility and versatility of EDAF, making it a valuable tool for wireless experimentation.

\end{document}